\documentclass[aps,prd,twocolumn,superscriptaddress]{revtex4-1}
\usepackage{graphicx}  
\usepackage{dcolumn}   
\usepackage{bm}        
\usepackage{amssymb}   
\usepackage{amsmath}
\usepackage{subcaption}
\usepackage{hyperref}
\usepackage{xcolor}
\usepackage{mathtools}

\DeclarePairedDelimiter\abs{\lvert}{\rvert}

\newcommand{\orcidicon}[1]{\href{https://orcid.org/#1}{\includegraphics[height=\fontcharht\font`\B]{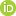}}}

\begin{document}

\title{Non-Kerr Constraints using Binary Black Hole inspirals considering phase modifications up to 4 PN order}

\author{Debtroy Das}
\affiliation{Center for Astronomy and Astrophysics, Center for Field Theory and Particle Physics, and Department of Physics,\\
Fudan University, Shanghai 200438, China}

\author{Swarnim Shashank~\orcidicon{0000-0003-3402-7212}}
\affiliation{Center for Astronomy and Astrophysics, Center for Field Theory and Particle Physics, and Department of Physics,\\
Fudan University, Shanghai 200438, China}

\author{Cosimo Bambi~\orcidicon{0000-0002-3180-9502}}
\email[Corresponding author: ]{bambi@fudan.edu.cn}
\affiliation{Center for Astronomy and Astrophysics, Center for Field Theory and Particle Physics, and Department of Physics,\\
Fudan University, Shanghai 200438, China}
\affiliation{School of Natural Sciences and Humanities, New Uzbekistan University, Tashkent 100007, Uzbekistan}

\date{\today}

\begin{abstract}
The gravitational field around an astrophysical black hole (BH) is thought to be described by the Kerr spacetime, which is a solution of the Einstein equation. Signatures of binary black hole (BBH) coalescence in gravitational waves (GW) follow the Kerr spacetime as the theoretical foundation. Hence, any possible deviations from the Kerr spacetime around BHs serve as a test of the nature of gravity in the strong-field regime and of the predictions of General Relativity. In our study, we perform a theory-agnostic test of the Kerr hypothesis using BBH inspirals from the third Gravitational-wave Transient Catalog (GWTC-3). Considering the Johannsen metric, we compute the leading-order deviation to the emitted GW in the frequency domain. Our results provide constraints on two deformation parameters ($\alpha_{13}$ and $\epsilon_3$) and demonstrate the degeneracy between these two non-Kerr parameters.
\end{abstract}

\maketitle

\section{Introduction \label{sec:1}}

Einstein’s Theory of General Relativity (GR), published in 1916, led to the prediction of the existence of gravitational waves (GW): ripples in spacetime generated by highly energetic astrophysical processes that would be so faint and their interaction with matter so weak that Einstein himself wondered if they could ever be discovered. However, scientists have tried to detect them using various methods. Finally, 100 years after the theory of GR, on 14 September 2015, these waves were directly detected by the Laser Interferometer GW Observatories~\cite{abbott2016gw150914}. The international scientific collaboration of LIGO-Virgo-KAGRA (LVK) conducted three observation runs and listed several events \cite{abbott2019gwtc,abbott2021gwtc,abbott2024gwtc,abbott2023gwtc}. These events are transient GW chirp signals from the inspiral and merger of compact binaries, like binary black holes (BBH) and binary neutron star (BNS) systems. The Observing Run 4 (O4) is ongoing and shall record data for 18 months \cite{ligoIGWNObserving}.

GW astronomy has opened a new frontier in astronomy. GW carry information on the motions of compact objects, and their weak interaction with matter allows us to see greater distances in space. Hence, GW are a new window through which we probe the universe and is the state-of-the-art tool to test the theory of GR in the dynamical and strong gravity regime. The theory has been subjected to several tests in past decades, mainly in the weak gravity regime~\cite{Will:2014kxa}, where most observational tests comply with theoretical predictions.

For the strong gravity regime, testing the Kerr hypothesis \cite{bambi2017testing,konoplya2016general,ghasemi2016note,mazza2021novel} in the electromagnetic spectrum has been attempted by using X-ray data from NuSTAR, RXTE, Suzaku, and XMM-Newton \cite{cao2018testing,tripathi2019toward,tripathi2020testing,tripathi2021testing,tripathi2021testing1,zhang2022testing} and radio data from the Event Horizon Telescope (EHT) experiment \cite{psaltis2021probing,bambi2019testing,psaltis2020gravitational,volkel2021eht}. In the latest “Tests of GR with GWTC-3” \cite{abbott2021tests}, LVK summarizes nine different analyses that use GW observations to test GR. These include consistency checks of the signal, tests of the GW generation, propagation, and properties like polarizations, and testing the BH ringdown hypothesis to probe the nature of the remnant object. Consistency tests refer to the Inspiral-Merger-Ringdown consistency test \cite{healy2017remnant,hofmann2016final,johnson2016determining} and residual (noise in the detector network after removing the GW signal from the data) analysis, indicating consistency between the signal in the data and the GR template used \cite{abbott2019tests}. The most significant tests related to the GW generation are parameterized tests of GR for generic modifications \cite{Yunes_Yagi_Pretorius_2016,nair2019fundamental} and tests with spin-induced quadrupole moment \cite{saleem2022population,krishnendu2019constraints}. Other tests involve searches of non-GR polarizations \cite{wong2021null}, non-GR dispersion of GW \cite{mirshekari2012constraining}, and signatures for GW echoes \cite{abedi2017echoes,tsang2018morphology,lo2019template}.

From a theoretical perspective, tests of GR can be categorized into two strategies, which are usually referred to as the theory-specific (or top-down) approach and the theory-agnostic (or bottom-up) approach. In the theory-specific approach, one way to test gravity is by analyzing GW signals with a GR model and a specific beyond GR gravity theory model and checking which of the two models can explain the data better. Another way is to constrain some parameters in the gravity theory beyond GR, which often reduces to GR in some limit. We do not want to test any specific gravity theory beyond GR in the theory-agnostic approach. We look for possible signatures of deviations from the predictions of GR.

In the present study, we use a parametrized metric to measure possible deviations from the Kerr solution in the approach similar to \cite{bambi2021towards,shashank2022constraining}. We consider the Kerr-like metric proposed by Johannsen, which depends on a set of free parameters along with the mass and the spin \cite{johannsen2013regular}. This metric reduces smoothly to the Kerr metric if all the additional parameters vanish. Hence, we follow the theory-agnostic approach. We aim to constrain the Johannsen metric's deformation parameters for stellar-mass BBH systems using GW data from the GWTC-3 catalog published by the LVK collaboration~\cite{abbott2023gwtc}. We focus on the inspiral part of the BH coalescence since that can be well-approximated through analytic approaches like the post-Newtonian (PN) formalism \cite{blanchet1989post}. Our study further investigates the correlations between the posterior distributions of the deformation parameters. Besides generating robust constraints, the major benefit of our theory-agnostic approach to constrain the parameters for regular, well-behaved metrics that deviate
from the GR predictions is the direct comparison with the results from other tests of the gravity theory.

The manuscript is compiled as follows. Section~\ref{sec:2} constructs the mathematical formalism of our approach and the testing method. The first part of the section emphasizes the Johannsen metric, our model for the Kerr hypothesis testing, and the analytical expressions for the deformation terms in the GW phase using PN expansion. The later part describes the Bayesian inference and the computational tool employed for our parameter estimation. In Section~\ref{sec:3}, the selection criteria of GW events and the results of our analysis are put forth. We conclude the paper with a discussion in Section~\ref{sec:4}.

\section{Model and Methods \label{sec:2}}

\subsection{Metric and Waveform Modification \label{sec:2a}}

In this paper, we employ the Johannsen metric \cite{johannsen2013regular}, a Kerr-like BH metric that is regular outside the event horizon and possesses three independent constants of motion. The metric has a nonlinear dependence on four free functions to parameterize the potential deviations from the Kerr metric. In Boyer-Lindquist-like coordinates, if we consider equatorial circular orbits ($\theta = \pi/2$ and $\dot{\theta} = 0$), the metric is expressed by the given non-vanishing components of the line element \cite{johannsen2013regular}
\begin{eqnarray}
\begin{aligned}
    &g_{tt} = - \frac{\Tilde{\Sigma}[\Delta - a^2 A_2(r)^2]}{[(r^2 +a^2)A_1(r) - a^2 A_2(r)]^2} \\
    &g_{t\phi} = - \frac{\Tilde{\Sigma} a [(r^2 +a^2)A_1(r) A_2(r) - \Delta]}{[(r^2 +a^2)A_1(r) - a^2 A_2(r)]^2} \\
    &g_{rr} = \frac{\Tilde{\Sigma}}{\Delta A_5(r)} \\
    &g_{\theta \theta} = \Tilde{\Sigma}\\
    &g_{\phi \phi} = \frac{\Tilde{\Sigma} [(r^2 +a^2)^2 A_1(r)^2 - a^2 \Delta]}{[(r^2 +a^2)A_1(r) - a^2 A_2(r)]^2}
\end{aligned}
\end{eqnarray}
where,
\begin{equation}
    A_1(r) = 1 + \sum_{n=3}^{\infty} \alpha_{1n} \left(\frac{M}{r} \right)^n
\end{equation}
\begin{equation}
    A_2(r) = 1 + \sum_{n=2}^{\infty} \alpha_{2n} \left(\frac{M}{r} \right)^n
\end{equation}
\begin{equation}
    A_5(r) = 1 + \sum_{n=2}^{\infty} \alpha_{5n} \left(\frac{M}{r} \right)^n
\end{equation}
\begin{equation}
    \Tilde{\Sigma} = r^2 + f(r)
\end{equation}
\begin{equation}
    f(r) = \sum_{n=3}^{\infty} \epsilon_n \frac{M^n}{r^{n-2}}
\end{equation}
This version of the line element already meets the Newtonian limit and has no constraints from weak field experiments in the Solar System \cite{johannsen2013regular}. The leading order non-Kerr parameters are $\alpha_{13}$, $\alpha_{22}$, $\alpha_{52}$, and $\epsilon_3$. In our work, for the sake of simplicity, we only consider the possibility of non-vanishing $\alpha_{13}$ and $\epsilon_3$ (and, as we show below, we can only constrain one of them when we assume that the other one vanishes).

To obtain the GW phase, we follow the parametrized post-Einsteinian (ppE) formalism explored in Refs.~ \cite{shashank2022constraining,cardenas2020gravitational,Yunes_Yagi_Pretorius_2016,Yunes_Pretorius_2009}. First, we obtain the effective potential ($V_{\rm eff} = g_{rr}\dot{r}^2$) using the four-velocity normalization $u^{\mu} u_{\mu} = -1$ which gives
\begin{equation}
    V_{\rm eff} = - g_{\phi \phi}\dot{\phi}^2 - g_{tt}\dot{t}^2 - 1 \, ,
\end{equation}
where $V_{\rm eff} = V_{\rm eff}^{\rm GR} + V_{\rm eff}^{\rm NK}$. Here and in the following equations, super-script \emph{NK} refers to the terms relating to the deformation parameters of the metric, which provide deviation from GR. Solving for $V_{\rm eff} = d V_{\rm eff}/dr = 0$, we can obtain the deviation in the energy and angular momentum of circular orbits
\begin{eqnarray}
    E = E^{\rm GR} + E^{\rm NK} \, , \\
    L = L^{\rm GR} + L^{\rm NK} \, .
\end{eqnarray}
In the far-field limit, $L = r^2 \Omega$, where $\Omega$ is the angular velocity measured by a distant observer. This provides a modified Kepler's law, which reads
\begin{equation}\label{eq:kepler}
    \Omega^2 = \frac{M}{r^3} \left[ 1 + \frac{3 M}{r} + \frac{9 M^2}{r^2}+\frac{81 M^3}{8 r^3} +\frac{729 M^4}{64 r^4}  \right] + (\Omega^{\rm NK})^2 \, .
\end{equation}
From here, using the methods described in Refs.~[\cite{shashank2022constraining,cardenas2020gravitational,Buonanno:1998gg,Hinderer:2017jcs}], we can find a correction in the binding energy
\begin{equation}
    E_b = E_b^{\rm GR} + E_b^{\rm NK} \, .
\end{equation}
Now, the orbital phase is given as,
\begin{equation}
    \phi(\nu) = \int^{\nu} \frac{1}{\dot{E}} \left(\frac{dE}{d \Omega}\right)  \Omega ~ d \Omega
\end{equation}
where $\dot{E}$ is the energy dissipated by gravitational waves. We assume dissipative effects to be the same as GR, so we only take the 0PN term \cite{shashank2022constraining,cardenas2020gravitational,nissanke2005gravitational}. Finally, we obtain the orbital phase as 
\begin{equation}
    \phi(\nu) = \phi^{\mathrm{GR}}(\nu) + \phi^{\rm NK}(\nu) \, .
\end{equation}
Taking the Fourier transform of $\phi$ in the stationary phase approximation \cite{shashank2022constraining,cardenas2020gravitational} we get the GW phase
\begin{equation} \label{eq:phase_gr_nk}
    \Psi_{\mathrm{GW}}(f)=\Psi_{\mathrm{GW}}^{\mathrm{GR}}(f) + \Psi_{\mathrm{GW}}^{\rm NK}(f) \, .
\end{equation}
This $\Psi_{\mathrm{GW}}^{\rm NK}$ is the term we add in the GW waveform model and carry out our analysis using \texttt{bilby}~\cite{Ashton:2018jfp} with an extra parameter (deformation) to be estimated.

From Ref.~\cite{Carson:2020iik}, we see that the PN order in the ppE phase can be estimated by the powers of $M$ in Eq.~(\ref{eq:kepler}). In this work, we keep the terms up to the 4PN order. We did not calculate any coupling arising between spin and deformation in the metric. In the waveform model as well, we keep all the GR terms up to the 4PN order in phase to do our fitting. The non-vanishing GW phases of the deformation terms in the Johannesen metric up to the 4 PN order are as follows. For $\alpha_{13}$, the additional GW phase because of the deformation is 
\begin{equation}
\begin{split}
\Psi_{\mathrm{GW}}^{\rm NK} = -\frac{45 \alpha_{13} \kappa \sqrt[3]{u}}{4 \eta^{6/5}}+\frac{10 \alpha_{13} \kappa u}{3 \eta ^{8/5}}-\frac{15 \alpha_{13} \kappa }{4 \eta ^{4/5}\sqrt[3]{u}}
\\
-\frac{10 \alpha_{13} \kappa u \log (u)}{3 \eta ^{8/5}} \, . 
\end{split}
\end{equation}
For $\epsilon_{3}$, the additional GW phase because of the deformation is 
\begin{equation}
\begin{split}
\Psi_{\mathrm{GW}}^{\rm NK} = \frac{45 \epsilon_{3} \kappa \sqrt[3]{u}}{8 \eta ^{6/5}}-\frac{5 \epsilon_{3}\kappa u}{3 \eta ^{8/5}}
+\frac{15 \epsilon_{3} \kappa }{32 \eta ^{4/5} \sqrt[3]{u}}
\\
+\frac{5 \epsilon_{3} \kappa u \log (u)}{3 \eta ^{8/5}} \, . 
\end{split}
\end{equation}
We only consider modifications in the phase in our paper since the ppE formalism accounts for only phase modifications. While waveform modeling, we consider the amplitude at 0PN. In the literature, there are studies where both phase and amplitude modification are considered to probe BBH inspirals (see Ref. \cite{psaltis2021probing}).

\begin{figure}[h]
	\centering
	\begin{subfigure}{\linewidth}
		\includegraphics[scale=0.4]{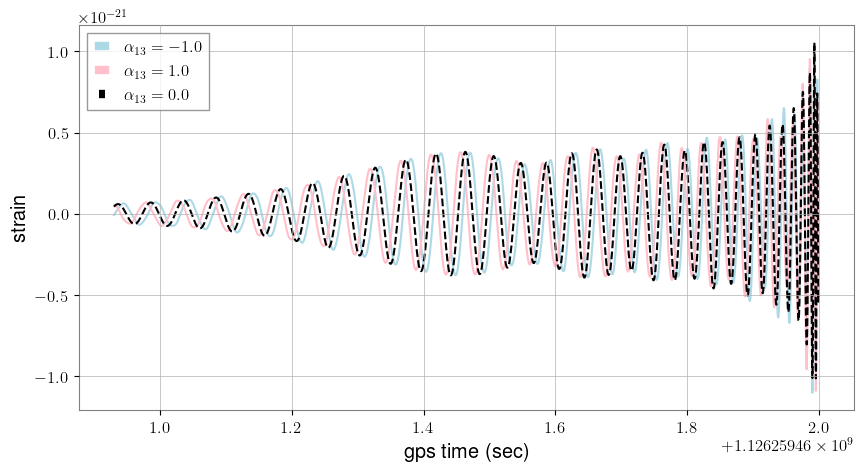} \\
		\includegraphics[scale=0.4]{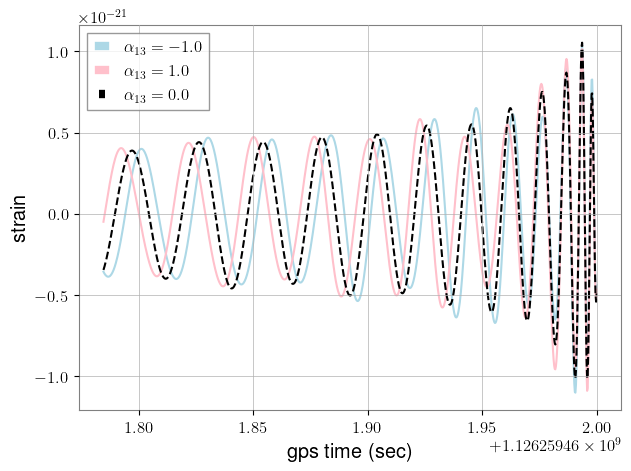}
	\end{subfigure}
         \vfill
	\caption{Modified waveform model showing the effect of $\alpha_{13}$ for $\epsilon_3 = 0$. The bottom panel is a zoom of the last part of the waveform in the top panel.}
	\label{fig:1}
\end{figure}

\begin{figure}[h]
	\centering
	\begin{subfigure}{\linewidth}
		\includegraphics[scale=0.4]{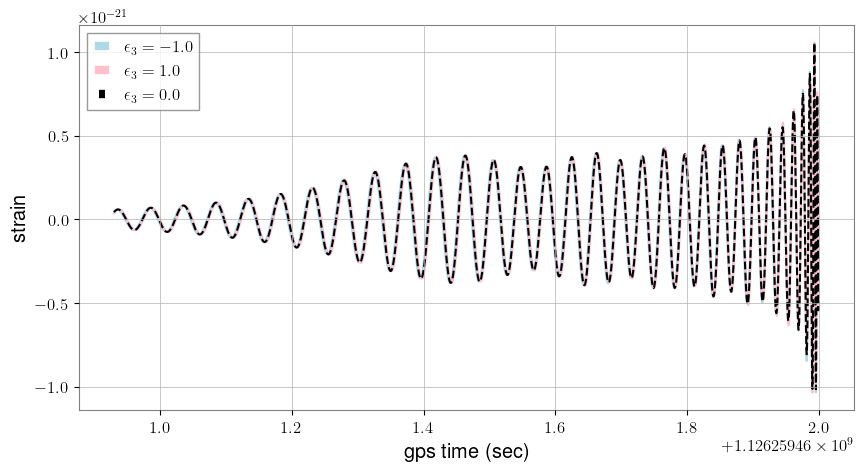} \\
		\includegraphics[scale=0.4]{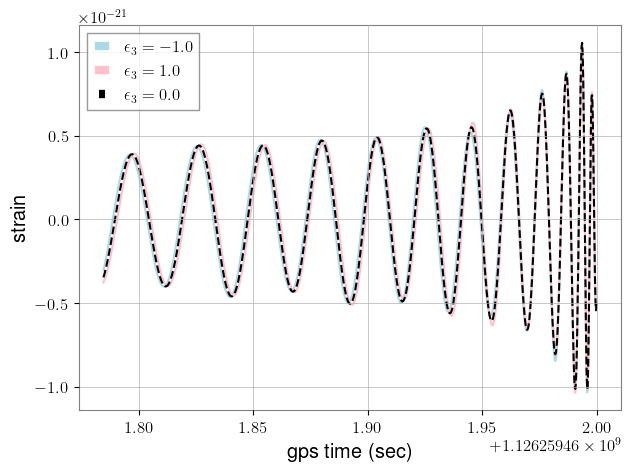}
	\end{subfigure}
	\caption{Modified waveform model showing the effect of $\epsilon_3$ for $\alpha_{13} = 0$. The bottom panel is a zoom of the last part of the waveform in the top panel.}
	\label{fig:2}
\end{figure}

To study the deviations from the Kerr geometry from GW data, we employ the precessing aligned-spin waveform model~\verb+IMRPhenomP_v2+~\cite{Husa:2015iqa, Khan:2015jqa} incorporating the additional phase term arising due to the deviation from Kerr metric, as shown in Eq.~(\ref{eq:phase_gr_nk}). Fig.~\ref{fig:1} and \ref{fig:2} demonstrate the modified ~\verb+IMRPhenomP_v2+ waveform considering positive and negative values of one of the non-Kerr parameters while the other non-Kerr parameter is set to 0. We have explored the changes in the waveform for $\alpha_{13} = \pm 1$ and $\epsilon_3 = 0$ in Fig.~\ref{fig:1}, and in the waveform for $\epsilon_3 = \pm 1$ and $\alpha_{13} = 0$ in Fig.~\ref{fig:2}.

\subsection{Analysis Framework \label{sec:2b}}

The waveform thus modeled are used to infer the BBH parameters along with the non-GR deformation parameters using Bayesian parameter estimation
\begin{equation} \label{eq:bayesian}
    p (\bm{\theta} \mid d) \propto \mathcal{L}(d \mid \bm{\theta}) \pi_{\rm PE} (\bm{\theta}) \,,
\end{equation}
where $\bm{\theta}$ includes all the source parameters and non-GR deformation parameters together and $\pi_{\rm PE}$ represents the prior used in inferring $\bm{\theta}$. In Eq.~\eqref{eq:bayesian}, $\mathcal{L}(d \mid \theta)$ corresponds to the likelihood of observing the GW signal $d$ given the parameters $\bm{\theta}$~\cite{Veitch:2009hd}; i.e.,
\begin{equation}
    \mathcal{L}(d \mid \bm{\theta}) \propto \exp \left(-\sum_{k} \frac{2\abs{d_{k}-h_{k}(\bm{\theta})}^{2}}{TS_{k}} \right) \,.
\end{equation}
Here, $h(\bm{\theta})$ denotes the gravitational waveform for the BBH signal with parameters $\bm{\theta}$, T is the duration of the data segment to be analyzed for inferring $\bm{\theta}$, $S_{k}$ is the noise power spectral density (PSD) characterizing the sensitivity of the detector, and $k$ signifies the frequency dependence of the data, waveform, and PSD.

We use the \verb+bilby+~\cite{Ashton:2018jfp} implementation of~\verb+dynesty+ sampler~\cite{2020MNRAS.493.3132S} to infer the posterior distribution of source and non-GR parameters by analyzing the real data corresponding to BBH mergers, detected by LIGO-Hanford and LIGO-Livingston detectors. We implement uniform priors for chirp mass and mass ratio for each event we consider in the study, in the same range as published in the latest version of the respective events by the LVK collaboration. The luminosity distance priors are power law priors in suitable ranges for each event from the feasible values as mentioned in Ref.~\cite{Ashton:2018jfp}. For the rest of the parameters, we consider the default priors implemented in~\verb+bilby+ (refer to Table $1$ in Ref.~\cite{Ashton:2018jfp} for the default priors). The priors of the non-Kerr parameters also have uniform distributions in the range $[-5,5]$. Considering values beyond this limit would imply large deviations from GR, which are not expected.

We use a conversion function to set theoretical constraints on the prior distribution for the non-GW parameters. The theoretically feasible bounds of $\alpha_{13}$ and $\epsilon_3$ are as follows \cite{johannsen2013regular}
\begin{eqnarray}
\begin{aligned}\label{eq-constarints}
    &\alpha_{13} > - \frac{(M + \sqrt{M^2 - a^2})^3}{M^3} \, , \\
    &\epsilon_3 > - \frac{(M + \sqrt{M^2 - a^2})^3}{M^3} \, , \\
\end{aligned}
\end{eqnarray}
where $M$ is the reduced mass of the binary system, defined as $M= m_1 m_2 / (m_1+m_2)$, where $m_1$ and $m_2$ are the individual BH masses and $a$ for our system of interest is defined as
\begin{equation}
    a = \frac{a_1 + (1/q)^2 a_2}{(1+1/q)^2} + \xi  \frac{1/q}{(1+1/q)^2} (a_1 + a_2) \, ,
\end{equation}
In the above equations, $a_1$ and $a_2$ are the dimensionless spin magnitude of the two BHs, and  $q$ is the mass ratio ($q= m_2/m_1$ and $m_1>m_2$), and $\xi = 0.474046$.

\section{Results \label{sec:3}}

The GW catalogs have listed 90 confident transient astrophysical events from merging binaries consisting of BHs and neutron stars. We have considered all the BBH events from GWTC-3 \cite{abbott2023gwtc,abbott2023open} and shortlisted 6 events, which are listed in Tab.~\ref{tab:1}. Our study's selection criteria consider events with a total mass of less than 80 $M_\odot$ and a network signal-to-noise ratio (SNR) greater than 15. Given our waveform model, analyzing events with negligible spin precession is only justified.  We use the $\chi_p$ values from the ~\verb+IMRPhenomXPHM+~posteriors from the LVK release to determine events with precessing aligned-spin\cite{pratten2021computationally,abbott2021gwtc,abbott2023gwtc}.Hence, we consider these three selection criteria for our study. We estimate the posterior distributions of the intrinsic and extrinsic BBH parameters along with the non-GR parameters using the Bayesian approach described in the previous section.

\begin{table}[b]
\caption{Summary of the constraints on the non-Kerr parameters $\alpha_{13}$ (assuming $\epsilon_3 = 0$) and $\epsilon_3$ (assuming $\alpha_{13}= 0$). When both non-Kerr parameters are free, none of them can be constrained. The error values correspond to 90\% confidence intervals. \label{tab:1}}
\begin{ruledtabular} 
\renewcommand\arraystretch{1.5}
\begin{tabular}{ccc} 
Event& $\alpha_{13}$\footnote{For $\epsilon_3 = 0$}& $\epsilon_3$\footnote{For $\alpha_{13}= 0$}  \\
\hline
GW150914& $-0.23^{+0.4}_{-0.39}$ & $1.02^{+1.64}_{-1.65}$ \\
GW170814 & $0.16^{+0.42}_{-0.42}$ & $-0.7^{+1.75}_{-1.75}$ \\
GW190630\_185205 & $0.11^{+0.7}_{-0.65}$ & $-0.56^{+2.73}_{-2.78}$  \\
GW190828\_063405 & $0.45^{+0.77}_{-0.72}$ & $-1.85^{+2.86}_{-2.53}$ \\
GW200112\_155838&$0.01^{+0.59}_{-0.51}$&$-0.01^{+1.97}_{-2.38}$ \\
GW200311\_115853 &$-0.36^{+0.98}_{-0.66}$&$1.51^{+2.58}_{-3.83}$ \\
\end{tabular}
\end{ruledtabular}
\end{table}

For each of the events, we run three parameter estimations: (i) where $\alpha_{13}$ is a free parameter and $\epsilon_3 = 0$; (ii) where $\epsilon_3 $ is a free parameter and $\alpha_{13}= 0$; (iii) both $\alpha_{13}$ and $\epsilon_3$ are free parameters. For every analysis run, we have allowed the non-Kerr parameters a parameter space in the range $[-5,5]$ as discussed in the analysis framework. The constraints on the non-GR parameters from the posterior distributions of our analysis comply with the theoretically feasible bounds on $\alpha_{13}$ and $\epsilon_3$ given in Eq.~(\ref{eq-constarints}).

\begin{figure*}
\begin{center}
\includegraphics[width = 0.8\paperwidth]{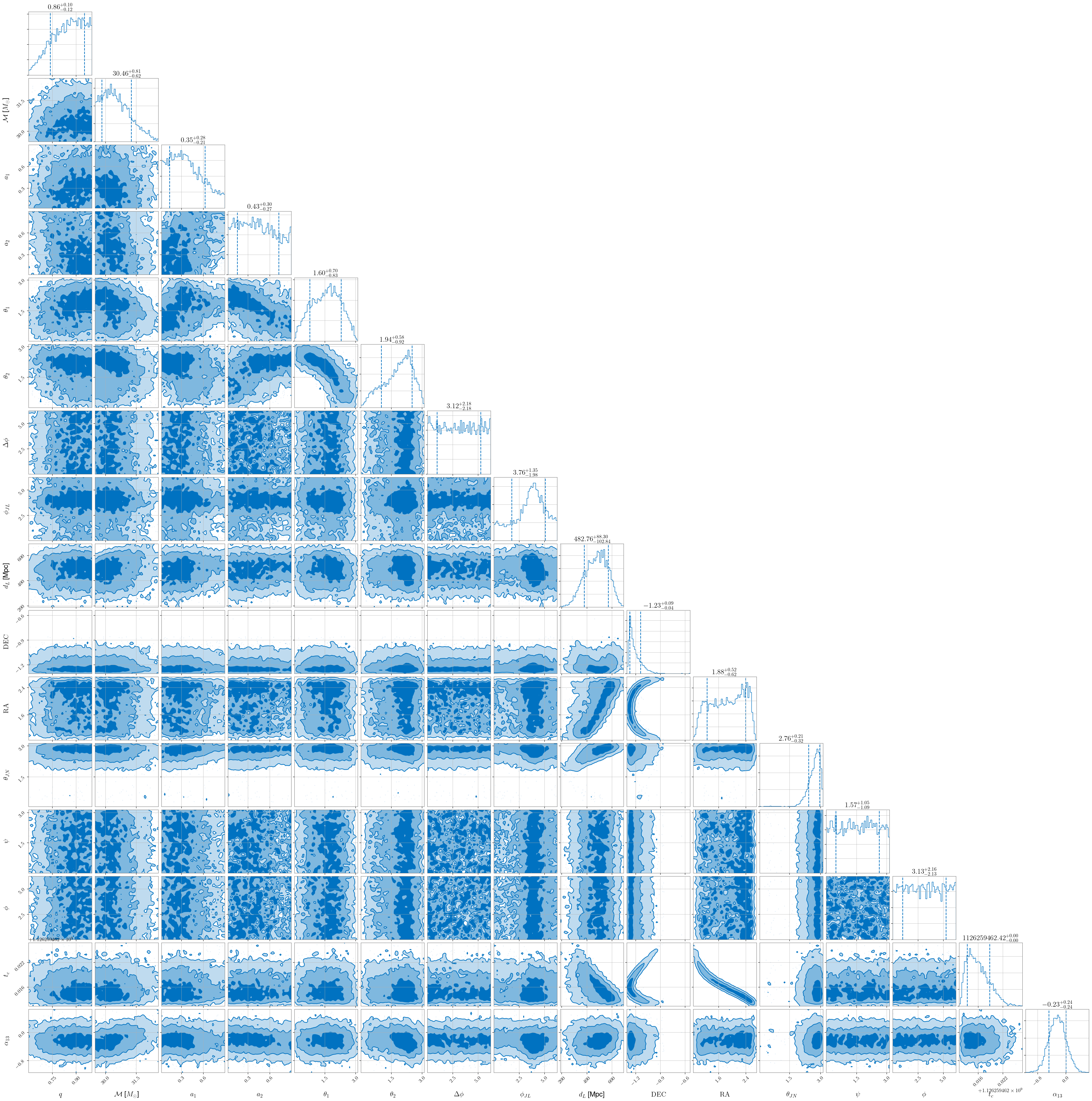}
\caption{\label{fig:epsart} Corner plot of the posterior distributions for GW150914 when $\alpha_{13}$ is free ($-5 \leq \alpha_{13} \leq 5$) and $\epsilon_3 =0$. \label{fig:3}}
\end{center}
\end{figure*}

\begin{figure*}
\begin{center}
\includegraphics[width = 0.8\paperwidth]{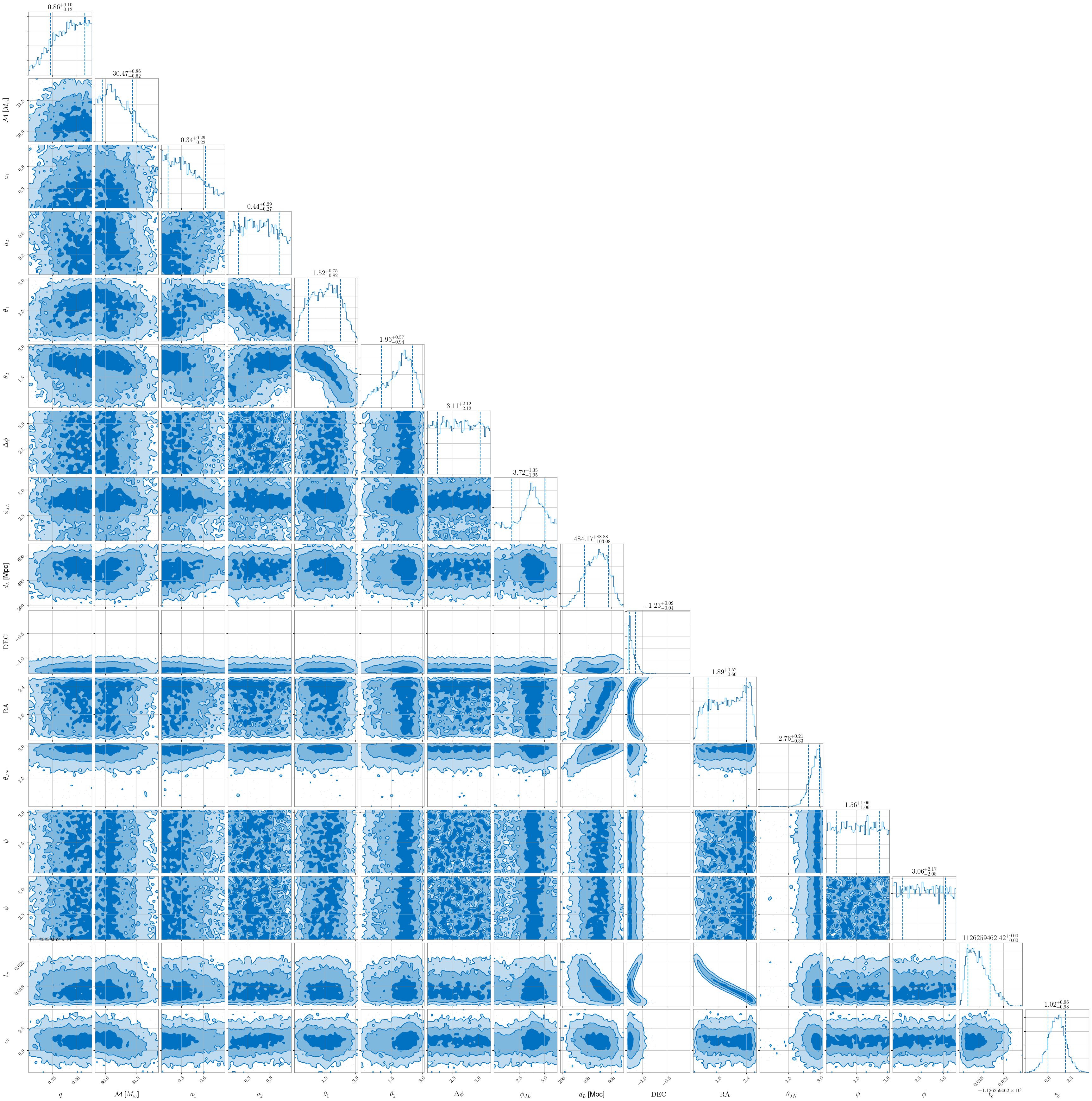}
\caption{\label{fig:epsart} Corner plot of the posterior distributions for GW150914 when $\epsilon_3$ is free ($-5 \leq \epsilon_3 \leq 5$) and $\alpha_{13} =0$. \label{fig:4}}
\end{center}
\end{figure*}

\begin{figure*}
\begin{center}
\includegraphics[width = 0.8\paperwidth]{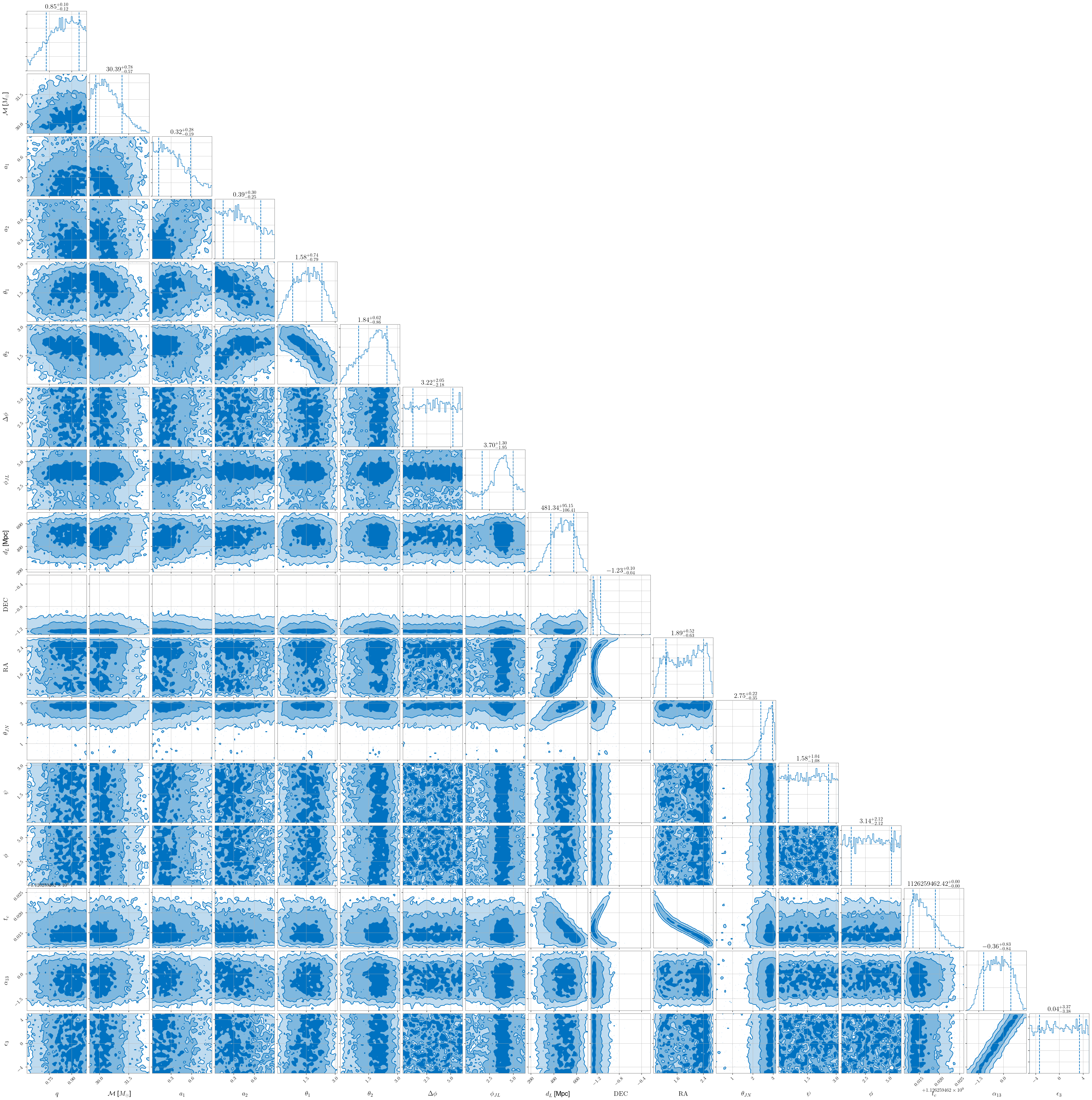}
\caption{\label{fig:epsart} Corner plot of the posterior distributions for GW150914 when both $\alpha_{13}$ and $\epsilon_3$ are free in the range $[-5,5]$. \label{fig:5}}
\end{center}
\end{figure*}

Figs.~\ref{fig:3}, \ref{fig:4} and \ref{fig:5} show the corner plots of the posteriors of the parameters of GW150914 obtained for the three cases as described above. The vertical dashed lines in the posterior distributions correspond to $1 \sigma$ confidence intervals of each parameter.  To examine our waveform model’s ability to retrieve GR,  we compare the 95 \% confidence intervals of parameters like chirp mass and luminosity distance between the ~\verb+IMRPhenomXPHM+~  posteriors from the LVK release and the posteriors obtained from our analysis\cite{pratten2021computationally,abbott2023open}.  We observe no significant deviations in these parameters.  The disparities in chirp mass and luminosity distance between the posteriors from our analyses and the GR predictions from the GWOSC \cite{abbott2023open} database for GW150914 and the other events are presented in the Suplementary material.  Furthermore, we notice the deformation parameters are constrained and statistically compatible with GR
at a high level of significance when one of them is free and the other is frozen to 0. When we have both the deformation parameters free, their value cannot be constrained. Although $\alpha_{13}$ appears to be constrained, it is just the result of the fact that $\epsilon_3$ is allowed to vary in the range $[-5,5]$ and the upper and lower constraints on $\alpha_{13}$ correspond to the case $\epsilon_3 = -5$ and 5, respectively. The estimates of these two parameters are highly correlated.

We run similar analyses for all the events shortlisted for our study. The results obtained from the other events are similar in nature. When one of the non-Kerr parameters is free and the other is frozen to 0, the free non-Kerr parameter is constrained around 0 and is consistent with the GR prediction. With both non-Kerr parameters free for the analysis, none of them can be constrained. The corner plots of the posterior distributions of these other five events are shown in the Supplementary Material.

\begin{figure}[h]
	\centering
	\begin{subfigure}{\linewidth}
		\includegraphics[scale=0.45]{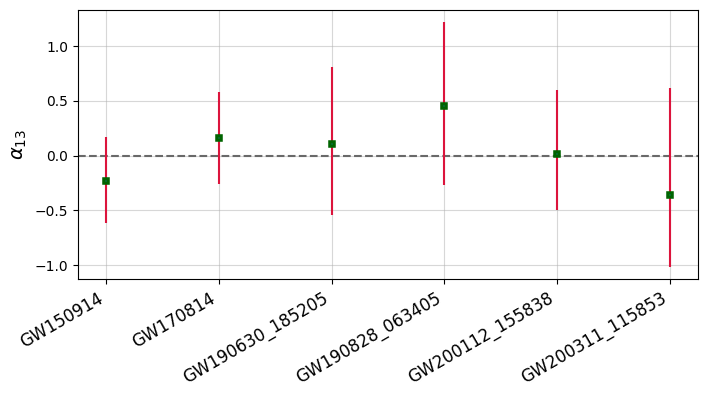}
		\caption{when $\alpha_{13}$ is free parameter and $\epsilon_3 = 0$}
	\end{subfigure}\\ \vspace{0.4cm}
	\begin{subfigure}{\linewidth}
		\includegraphics[scale=0.45]{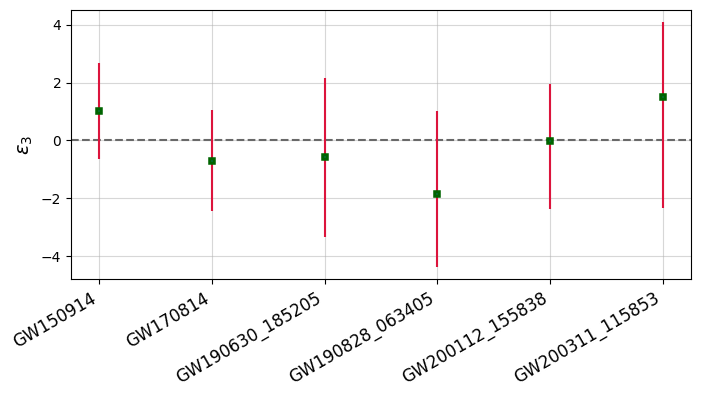}
		\caption{when $\epsilon_3$ is free parameter and $ \alpha_{13}= 0$}
	\end{subfigure}
	\caption{90\% confidence interval constraints on $\alpha_{13}$ (assuming $\epsilon_3 = 0$, top panel) and $\epsilon_3$ (assuming $\alpha_{13}= 0$, bottom panel) from the 6~events in GWTC-3 meeting our selection criteria. \label{fig:6}} 
\end{figure}

The 90\% confidence intervals of the constrained deformation parameters are shown in Tab.~\ref{tab:1} and Fig.~\ref{fig:6}.

\section{Discussion \label{sec:4}}

In this paper, we have investigated GW data of BBH inspirals, taking into account the Johanssen metric to test the Kerr hypothesis. We have limited our study to the deformation parameters $\alpha_{13}$ and $\epsilon_3$. If we assume that one of the two deformation parameters vanishes, we can constrain the other one. Unfortunately, our analysis is unable to constrain both $\alpha_{13}$ and $\epsilon_3$ simultaneously because their estimate is highly correlated. Our results are consistent with the Kerr hypothesis and the predictions of GR.

The theoretical framework of this paper is an extension of Ref.~\cite{shashank2022constraining}. The constraints reported in our current work are more accurate (as we have repeated the data analysis rather than using the posterior samples released by the LVK collaboration and we have included BH spins) and they turn out to be somewhat more stringent than those in \cite{shashank2022constraining} (if we compare the results of the events in GWTC-2, see Tab.~IV in \cite{shashank2022constraining}). As in Ref.~\cite{shashank2022constraining}, in our analysis we assume that the deformation parameters of the two BHs in the binary system have the same value. Relaxing this assumption would require significant changes in the derivation of the GW phase, which is out of the scope of this work and may be explored in the future.

In general, we can expect two main scenarios in models beyond GR violating the Kerr hypothesis. In the first scenario, BHs are not described by the Kerr solution, but some form of the no-hair theorem still holds and a BH is still completely specified by its mass and spin angular momentum. In such a case, the values of the deformation parameters should be the same for all BHs. In the second scenario, the no-hair theorem is violated and BHs are characterized by other parameters beyond the mass and the spin angular momentum. In such a case, the values of the deformation parameters would be, in general, different for different objects. There is also the possibility of something between these two main scenarios, when a BH has ``secondary hairs''. In this third case, there are other parameters characterizing a BH, but they are not independent and are instead determined by the values of the BH mass and/or spin. The Johannsen metric is not a BH solution of a specific theory of gravity but rather a parametric BH spacetime to test the Kerr hypothesis with astrophysical observations, so it cannot address the question of whether the values of $\alpha_{13}$ and $\epsilon_3$ should be the same for all BHs or not. If we assume that the values of $\alpha_{13}$ and $\epsilon_3$ are the same for all BHs, we could combine the constraints of all events to increase the SNR and get stronger constraints (assuming that single-event constraints are dominated by statistical uncertainties, which is likely the case for our shortlisted events).

The constraints on $\alpha_{13}$ and $\epsilon_3$ found in this work can be compared to the constraints on these parameters inferred by other studies and with other techniques. For example, Ref.~\cite{zhang2022testing} reports the $3\sigma$-constraints on $\alpha_{13}$ from stellar-mass BHs with X-ray data (their Tab.~5) and Ref.~\cite{Bambi:2022dtw} shows the most stringent $3\sigma$-constraints on $\alpha_{13}$ from stellar-mass and supermassive BHs with different techniques (Fig.~13). To simplify the comparison, we report the $3\sigma$-constraints from our 6~events in Tab.~\ref{tab:2}. In the case of X-ray tests, there are a few sources that provide very stringent constraints (e.g., GX~339--4 with $\alpha_{13} = -0.02_{-0.13}^{+0.03}$ at $3\sigma$). Constraints on $\epsilon_3$ from supermassive BHs with X-ray data have been reported in Ref.~\cite{Tripathi:2019fms} and they appear more stringent than those found here with GW data (and we can expect that the analysis of X-ray data from stellar-mass BHs can provide even more stringent constraints, as they are brighter sources). However, X-ray tests can unlikely improve appreciably in the next years with current X-ray observatories and we will need to wait for the next generation of X-ray telescopes: eXTP may be launched in 2028-2029 and Athena may be launched in the second part of the 2030s. GW tests of the Kerr hypothesis can probably improve faster.

\vspace{0.3cm}

{\bf Note added.} At the end of the preparation of this paper, we noted that a similar study was reported in Ref.~\cite{santos2024testing}, where the authors constrain the Johannsen-Psaltis metric~\cite{Johannsen:2011dh} using data from GWTC-3. Their metric is different, but the method is very similar. In our case, we tried to constrain even two deformation parameters at the same time (even if without success).

\begin{table}[h]
\caption{Summary of the $3\sigma$-constraints on the non-Kerr parameters from the analysis of this work. \label{tab:2}}
\begin{ruledtabular}
\renewcommand\arraystretch{1.5} 
\begin{tabular}{ccc} 
Event & $\alpha_{13}$\footnote{For $\epsilon_3 = 0$}& $\epsilon_3$\footnote{For $ \alpha_{13}= 0$} \\
\hline
GW150914& $-0.23 ^{+ 0.6 }_{- 0.62 }$ & $1.02 ^{+ 2.64 }_{- 2.67 }$ \\
GW170814 & $0.16 ^{+ 0.65 }_{- 0.63 }$ & $-0.7 ^{+ 2.74 }_{- 2.68 }$ \\
GW190630\_185205 & $0.11 ^{+ 1.17 }_{- 0.95 }$ & $-0.56 ^{+ 3.96 }_{- 4.11 }$ \\
GW190828\_063405 & $0.45 ^{+ 1.22 }_{- 1.13 }$ & $-1.85 ^{+ 4.34 }_{- 3.06 }$ \\
GW200112\_155838&$0.01 ^{+ 1.04 }_{- 0.81 }$&$-0.01 ^{+ 3.19 }_{- 3.98 }$ \\
GW200311\_115853 &$-0.36 ^{+ 1.59 }_{- 1.0 }$ & $1.51 ^{+ 3.36 }_{- 5.89 }$ \\
\end{tabular}
\end{ruledtabular}
\end{table}

\section*{Acknowledgement}
The authors would like to thank Tathagata Ghosh and Bikram Pradhan
 from Inter-University Centre for Astronomy and Astrophysics,  India for their valuable inputs and discussions throughout the project.
This work was supported by the National Natural Science Foundation of China (NSFC), Grant No.~12250610185 and 12261131497, and the Natural Science Foundation of Shanghai, Grant No.~22ZR1403400. 
D.D. also acknowledges support from the China Scholarship Council (CSC), Grant No. 2022GXZ005434.

\bibliographystyle{apsrev4-1}
\bibliography{references}

\end{document}